\documentclass[final,5p,times,twocolumn]{elsarticle}

\usepackage{hyperref}
\usepackage{xcolor}

\journal{Journal of Colloid and Interface Science}

\newcommand{\nn}{\nonumber}
\newcommand{\veps}{\varepsilon}

\begin{document}

\begin{frontmatter}

\title{Bistable colloidal orientation in polar liquid near a charged wall}

\author{Yoav Tsori}
\address{Department of Chemical Engineering, Ben-Gurion University of the 
Negev, Israel.}

\begin{abstract}
We examine the translation and rotation of an uncharged spheroidal colloid in 
polar solvents (water) near a charged flat surface. We solve the nonlinear 
Poisson-Boltzmann 
equation outside of the colloid in two dimensions for all tilt angles $\theta$ 
with respect to the surface normal. The colloid's size is assumed to be 
comparable to the Debye's length and hence field gradients are essential. The 
Maxwell stress tensor, including the ideal gas pressure of ions, is integrated 
over the colloid's surface to give the total force and torque on the colloid. 
From the torque we calculate the effective angular potential $U_{\rm 
eff}(\theta)$. The classical behavior where the colloid tends to align in the 
direction perpendicular to the surface (parallel to the field, $\theta=0$) is 
retrieved at large colloid-surface distances or small surface potentials. We 
find a surprising transition whereby at small separations or large potentials 
the colloid aligns parallel to the surface ($\theta=90^\circ$). Moreover, this 
colloid orientation is amplified at a finite value of the aspect ratio. This 
transition may have important consequences to flow of colloidal suspensions or 
as a tool to switch layering of such suspensions near a surface.

\end{abstract}


\end{frontmatter}


{\bf Introduction}

Electric forces 
occur naturally and play a vital role in liquids, polymers, and biological 
matter \cite{rudi_book_2001,rudi_book_2014}. They can also 
be a convenient external tool to tune the structure of soft-matter systems. 
When a liquid dielectric droplet is placed in external electric field, it 
elongates along the field, and the elongation is proportional to 
the field 
squared \cite{okonski1953,mason1962}. In liquids and polymers the electric 
shear forces lead to various interfacial instabilities and order-order 
transitions 
\cite{russell_steiner2001a,russell_steiner2001b,russell_steiner2000,
russell_science1996,russell_steiner_nature_mater2003,russell_mm2004,
russel2003,russel2002,tsori_rmp2009,boker_mm_2002,boker_mm_2003,
boker_soft_matter_2007,tsori_apacs,tsori_mm_2002,tsori_mm_2003}. 

Microscopic particles in dilute suspensions are 
dominated by random thermal motion but order can arise due to 
specific interactions \cite{chaikin_bibette_nature_2008}, 
shape-dependent effective entropic interactions 
\cite{glotzer_acs_nano_2013,glotzer_angewandte_2013,glotzer_pnas_2014,
pine_cocis_2011}, chirality \cite{chaikin_science_2017} or external forces 
\cite{chaikin_lubensky_book}. Magnetic forces lead to chaining of magnetic 
multipoles \cite{pine_jacs_2012} and to fascinating structures in ferrofluids 
\cite{rosensweig_book}.

For a non-spherical solid particle in uniform field, classical works show that 
the torque depends on the three shape depolarization factors, and is 
proportional to $\sin(2\theta)$, where $\theta$ is the tilt angle of the 
particle with respect to its {\it single} lowest-energy orientation 
\cite{LL_electrodynamics}. If numerous such particles are suspended in an 
insulating solvent they tend to aggregate into filamentous structures, and 
these electrorheological liquids found several applications 
\cite{halsey_science_1992,gast_adv_coll_interf_sci_1989,ruzicka_book}.

In recent years considerable advances have been made in the synthesis and
preparation of elongated colloids 
\cite{yamada_chemcom_2003,basavaraj_soft-matter_2013} and complex
self-assembled morphologies were reported 
\cite{basavaraj_rsc_adv_2015,jerolmack_langmuir_2017}.
The ability to change colloidal orientation and to 
control colloidal clusters is important for achieving advanced optical, flow, 
and mechanical properties of suspensions.

In this work we look in details on a solid colloid suspended in 
a polar solvent 
(e.g. water) near a 
charged surface. The colloid's size $a$ is neither very large nor very small 
compared to the length-scale characterizing the electric field, Debye 
screening length $\lambda_D$. We focus on the often 
overlooked regime between electroosmosis of spherical or non-spherical objects, 
where typically $a\gg\lambda_D$ 
\cite{ramos_jcis_1999,bazant_squires_jfm_2004,bazant_electrophoresis_2011,
miloh_phys_fluids_2015,ramos_cocis_2016,yariv_phys_fluids_2005,
bazant_jfm_2006} , and uniform fields, where 
$a\ll\lambda_D$. When the colloid is near a charged surface and $a\simeq 
\lambda_D$, the simultaneous act of image charges and screening near the 
particle and the surface results a rich behavior.

Below we restrict ourselves to equilibrium and assume no hydrodynamics flow 
or electric currents. Recall first the classical forces that act on a 
solid macroscopic body submerged in a liquid in a gravitational field in the 
$y$ direction: regardless of the shape of the body, the total force
is in the $y$ direction, ${\bf F}=F\hat{y}$. This Archimedes force is 
proportional to the body's volume and is positive or 
negative depending on the buoyancy of the body relative to the 
liquid. In addition, the torque on the body vanishes, $\tau=0$. This is true 
because in the gravitational field the pressure varies linearly with $y$. 
However, we show that this does not hold in the case of a colloid near a 
charged surface. While a dielectrophoretic-like force in the $y$ 
direction can be expected, the torque is highly non trivial. In contrast to the 
classical case of uniform electric field outlined above, here there are two (in 
general more) competing energy minima, and the colloid's orientation 
can switch from the classical orientation into another one. The relative 
importance of the competing minima is dictated by the distance from the 
surface, the surface potential, the colloid size, the permittivities of the 
colloid and solvent, and the colloid shape.

\section{Model}

\begin{figure}[!th]
\begin{center}
\includegraphics[width=0.45\textwidth,bb=35 1 490 790,clip]{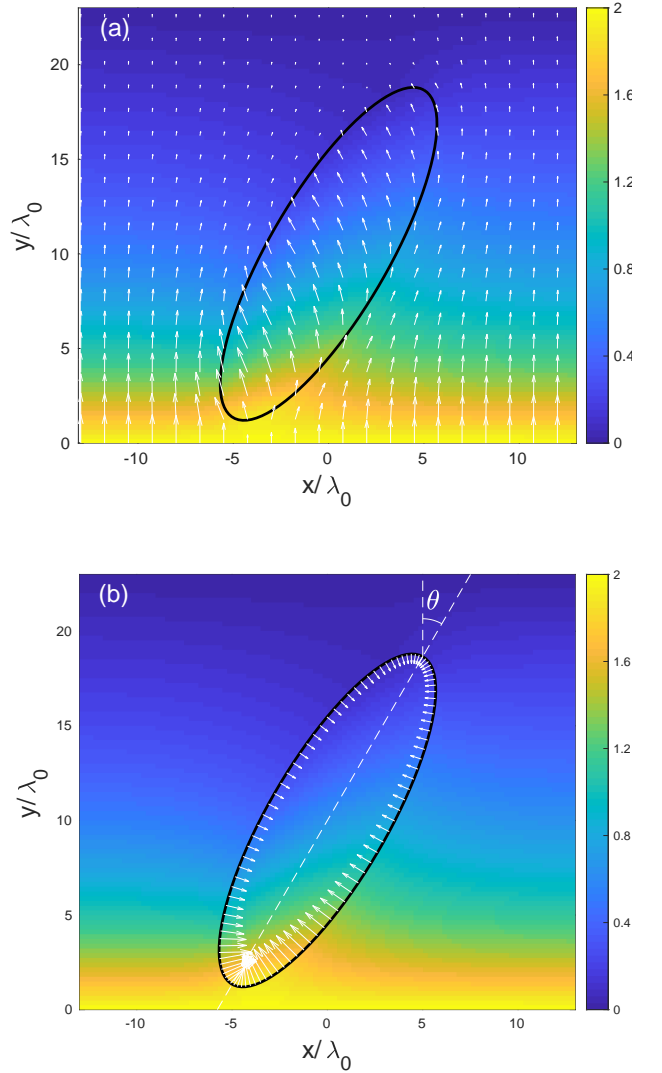}
\caption{(a) Numerical solution of Eq. (\ref{eq_pb}) for a dielectric colloid 
(black contour) in water near a charged wall. Arrows are proportional to the 
electric field while color is $\tilde{\psi}=e\psi/k_BT$.  The potential at the 
wall ($y=0$) is $\tilde{V}=2$. 
(b) $\theta$ is defined as the tilt angle between the colloid's long axis and 
the $y$-axis. Arrow show the surface force per unit area ${\bf f}_s$ acting on 
the colloid. Color is the potential as in (a). 
In this particular configuration $\theta=\pi/6$, $a=10\lambda_0$, 
$a/b=3.2$, and $y_{\rm 
center}=10\lambda_0$. In this and in other figures the 
dielectric constants are $\veps_c=2$ (colloid) and $\veps_w=80$ (solvent). 
(For interpretation of the references to colour in this figure legend, the 
reader is referred to the web version of this article.)
}
\label{fig1}
\end{center}
\end{figure}

The colloid is modeled as an uncharged solid ellipsoid of 
long and short axis $a$ and $b$, respectively. In the two-dimensional $x$-$y$ 
plane, this is a projection of long deformed cylinders extending in the 
$z$-direction.
The colloid, whose permittivity is $\veps_0\veps_c$, is placed in a polar 
solvent of permittivity $\veps_0\veps_w$ (e.g. water), with 
$\veps_0$ being the vacuum permittivity.
The charged surface is at $y=0$ and the colloid is close, the distance of its 
center of mass is $y=y_{\rm center}$. The colloid 
is tilted with respect to the surface such that its long axis makes an angle 
$\theta$ with the $y$ axis (see Fig. \ref{fig1} b). Our aim is to find the 
total force and torque acting on the colloid.

Within the mean-field theory, assuming point-like ions, and neglecting 
correlations, the electrostatic potential obeys the Poisson-Boltzmann equation 
\cite{levin_rep_prog_phys_2002,holm_pre_2004,levin_jcp_2009}:
\begin{eqnarray}
\veps_w\tilde{\nabla}^2\tilde{\psi}&=&\sinh(\tilde{\psi})~~~~{\rm 
outside~of~the~colloid}\nn\\
\veps_c\tilde{\nabla}^2\tilde{\psi}&=&0~~~~~~~~~~~~~~{\rm 
inside~the~colloid}\label{eq_pb}
\end{eqnarray}
The potential $\tilde{\psi}=e\psi/k_BT$ is scaled using the  
electron's charge $e$, the Boltzmann's constant $k_B$, and the 
absolute temperature $T$. Lengths are 
scaled as $\tilde{{\bf  r}}={\bf r}/\lambda_0$, where 
$\lambda_0$ is given by 
\begin{eqnarray}
\lambda_0^2=\frac{\veps_0 k_BT}{2n_0e^2}
\end{eqnarray}
Here $n_0$ is the bulk ion density far from the charged wall and colloid.
The Debye length is larger than $\lambda_0$; in water 
($\veps_w\approx 80$) $\lambda_D\approx 9\lambda_0$.
Equation (\ref{eq_pb}) obeys the boundary conditions 
$\tilde{\psi}(y=0)=\tilde{V}$, $\tilde{\psi}=0$ as $y\to\infty$, and 
sufficiently far from the colloid the field is oriented in the $y$-direction. 
The discontinuity of the normal field across the interface between the 
colloid and the solvent is obtained as $\llbracket {\bf 
D}\rrbracket\cdot\hat{n}=\sigma$, where ${\bf D}=\veps_0\veps{\bf E}$, 
and $\llbracket 
{\bf D}\rrbracket\equiv {\bf D}^{(2)}-{\bf D}^{(1)}$ is the jump in ${\bf D}$ 
across the regions. 
$\hat{n}$ is the normal unit vector pointing from region 1 (colloid) to region 
2 (solvent).
In this work the surface charge density $\sigma$ vanishes.
The continuity of the tangential field across the interface is given by 
$\llbracket {\bf 
E}\rrbracket\cdot\hat{t_i}=0$, where $\hat{t}_i$ ($i=1$, $2$) are the two 
orthogonal unit vectors lying in the plane of the interface. 

Inside the solid colloid, the nonuniform field leads to internal elastic 
stress. We assume large elastic modulii and the corresponding strain and energy 
are therefore vanishingly small. 
The stress tensor $\mathbb{T}$ is given by \cite{panofsky_phillips_book}:
\begin{eqnarray}\label{stress_tensor}
\mathbb{T}&=&-p_0(n^\pm,T)\delta_{ij}+\frac12\veps_0\veps 
E^2\left(-1+\rho(\partial\veps/\partial\rho)_T/\veps\right)\delta_{ij}\nn\\
&+&\veps_0\veps E_iE_j~.
\end{eqnarray}
$p_0$ includes the non-electrostatic contributions to $\mathbb{T}$ 
and depends on the density of the cations and anions $n^\pm$. In this work we 
take it as the ideal-gas pressure of the ions: $p_0=(n^++n^-)k_BT$. 
The second term, depending on the colloid's density $\rho$, includes 
electrostriction. Since it is diagonal in $\mathbb{T}$ it can be lumped together 
with $p_0$ without changing the forces on the colloid.
The body force in the liquid ${\bf f}$, given 
as a divergence of the stress: ${\bf f}_i=\partial\mathbb{T}_{ij}/\partial x_j$, 
and the force acting on a unit area of the interface between the colloid and the 
solvent, ${\bf f}_s$, are
\begin{eqnarray}\label{body_surface_force}
{\bf f}&=&-\nabla p_0+\frac12\nabla\left(\veps_0E^2 
\rho\frac{\partial \veps}{\partial 
\rho}\right)_T-\frac12\veps_0E^2\nabla\veps\nn\\
&+&e(n^+-n^-){\bf E}~,\nn\\
{\bf f}_s&=&\mathbb{T}\hat{n}~.
\end{eqnarray}

To reduce the computation time in the numerical procedure employed 
below, we convert the volume integrals to surface integrals in the following 
way. The volume integral for the force on the colloid ${\bf F}=\int_v{\bf 
\nabla}\cdot\mathbb{T}dv$ is converted to a surface integral 
$\int_S\mathbb{T}\hat{n}ds$ by virtue of the divergence 
theorem. For the torque, we use the Levi-Civita anti-symmetric 
tensor $\veps_{ijk}$ and write $\int_v\frac{\partial}{\partial 
r_l}\left(\veps_{ijk}r_j\mathbb{T}_{lk}\right)dv=\int_v\veps_{ijk}\mathbb{T}_{jk
}dv+\int_v\veps_{ijk}r_j\frac{\partial \mathbb{T}_{lk}}{\partial r_l}dv$. The 
term $\veps_{ijk}\mathbb{T}_{jk}$ vanishes due to the symmetry of $\mathbb{T}$ 
so the right-hand side is simply the i'th component of the torque 
$\tau=\int_v({\bf r}\times{\bf f})dv$. On the other hand, the divergence 
theorem gives 
$\int_v\frac{\partial}{\partial 
r_l}\left(\veps_{ijk}r_j\mathbb{T}_{lk}\right)dv=\int_s\veps_{ijk}r_j\mathbb{T}
_{lk}n_lds$, and this is the i'th component of $\int_s{\bf 
r}\times(\mathbb{T}\hat{n})ds$. Once the field and ion distributions are known 
we integrate these surface integrals to calculate the total force and 
torque acting on the colloid \cite{stratton_book}. 

When the torque $\tau(\theta,y_{\rm center})$ for given tilt angle 
$\theta$ and distance from the wall $y_{\rm center}$ is known, the effective 
rotation potential $U_{\rm eff}(\theta,y_{\rm center})$, defined as 
\begin{eqnarray}
\tau(\theta,y_{\rm center})=-\frac{U_{\rm eff}(\theta,y_{\rm 
center})}{d\theta}~,
\end{eqnarray}
can be calculated.

\section{Results}

We solve numerically Eq. (\ref{eq_pb}) with its boundary conditions on a 
rectangular grid in the $x$-$y$ plane, with a charged wall at $y=0$. Fig. 
\ref{fig1} (a) shows the potential (color) and field (arrows) distribution 
inside and near the colloid. In (b) the arrows are the surface force per 
unit area ${\bf f}_s$ acting on the colloid. Clearly, these forces are 
non-uniform, and they lead to a net translation and rotation. We 
look at a 
solvent that is more polar than the colloid and hence there is a trivial force 
in the $y$-direction pushing the colloid away from the surface. When the 
colloid is more polar than the solvent this force is reversed.
The system is invariant to reflections, namely when $\theta\to-\theta$ the 
energy stays the same while the torque reverses its sign. We therefore look at 
positive values of $\theta$. 

\begin{figure}[!th]
\begin{center}
\includegraphics[width=0.45\textwidth,bb=100 235 480 540,clip]{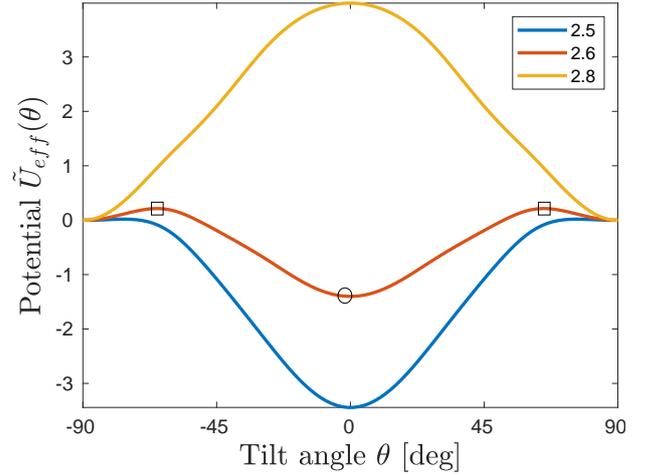}
\caption{Dimensionless effective rotational potential 
$\tilde{U}_{\rm eff}(\theta)=(4\pi l_{B,0}/L_z)U_{\rm eff}(\theta)/k_BT$ vs 
colloid tilt angle $\theta$. $l_{B,0}$ is the ``vacuum'' Bjerrum length given by 
$l_{B,0}\equiv e^2/(4\pi\veps_0k_BT)$ and $L_z$ is the length of the system in 
the $z$-direction. The three curves correspond to different values of surface 
potential $\tilde{V}=2.5$, $2.6$, and $2.8$. At sufficiently small voltage the 
minimum is at $\theta=0$, meaning that the colloid tends to orient parallel to 
the average field and perpendicular to the surface. 
At an intermediate voltage, $\tilde{V}=2.6$, $\theta=0$ is still a global 
minimum (circle) but $\theta=\pm90^\circ$ are local minima. Two global maxima 
develop at $\theta\approx\pm 63^\circ$ (rectangles). As the potential increases 
to $\tilde{V}=2.8$, the preferred colloidal orientation becomes {\it parallel} 
to the field, as the maximum is at $\theta=0$ and the minima are at $\theta=\pm 
90^\circ$. Note that since $L_z$ is expected to be much larger than 
$l_{B,0}\approx 5.6\times 10^{-8}$m, the depth of the energy minima of 
$U_{\rm eff}$ can be very large compared to the thermal energy $k_BT$. We used
$a=6\lambda_0$, $a/b=3.2$, and $y_{\rm center}=6.5\lambda_0$.
}
\label{fig2}
\end{center}
\end{figure}

In Fig. \ref{fig2} we plot the effective rotation potential vs tilt angle 
$\theta$ for a colloid that is allowed to freely rotate but its 
distance from the surface is fixed. For small potentials $\tilde{V}$ the 
minimum is, as expected, at $\theta=0$, namely the colloid tends to orient 
parallel to the average direction of the field ($y$ direction, perpendicular to 
the surface). As $\tilde{V}$ increases, $\theta=0$ stays the global minimum but 
local minima appear at $\theta=\pm90^\circ$ (colloid's long axis parallel to 
the surface). Further increase in $\tilde{V}$ decreases the minima at 
$\theta=\pm90^\circ$ until they become the global minima. At this point the 
preferred state is where the colloid lies parallel to the surface. The 
depth of the minima of $U_{\rm eff}/k_BT$ is of order $L_z/(4\pi l_{B,0})$, 
where $L_z$ is the colloid's typical size and $l_{B,0}\equiv 
e^2/(4\pi\veps_0k_BT)$ is the ``vacuum'' Bjerrum length. Since for a 
micron-sized colloid $L_z\sim 1$ $\mu$m and $l_{B,0}\approx5.6\times 10^{-8}$m 
we find that $U_{\rm eff}/k_BT\gg 1$.

\begin{figure}[!th]
\begin{center}
\includegraphics[width=0.45\textwidth,bb=95 240 480 540,clip]{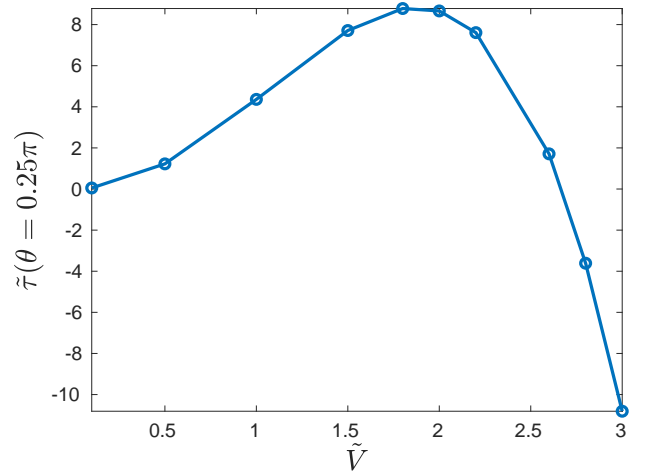}
\caption{Dimensionless torque acting on the colloid as defined by 
$L_z/(4\pi l_{B,0})\tilde{{\bf 
\tau}}={\bf \tau}/k_BT$, vs increasing values of surface potential $\tilde{V}$.
The tilt angle $\theta=\pi/4$ is fixed. The torque increases with $\tilde{V}$ to 
positive values, favoring counter clockwise (CCW) rotation. The torque levels 
off at $\tilde{V}\approx 2$ and decreases to negative values (clockwise 
rotation) 
at larger potentials. For these potentials the colloid's preferred orientation 
is parallel to the surface. $a=6\lambda_0$, $a/b=3.2$, and
$y_{\rm center}= 6.5\lambda_0$.
}
\label{fig3}
\end{center}
\end{figure}

To get a better understanding of this peculiar transition of the 
colloid from 
orientation parallel to an orientation perpendicular to the field, in Fig. 
\ref{fig3} we look at the total torque acting on the colloid with increasing 
potentials $\tilde{V}$ for a fixed orientation and distance from the surface. 
As 
$\tilde{V}$ increases the torque increases from zero. 
Positive values mean rotation in the CCW direction, tending to 
orient the colloid with its long axis parallel to the field. However, there is 
finite value of $\tilde{V}$ where the torque reaches a maximum, further 
increase in the potential leads to a {\it reduction} of the torque until, at 
another finite value, the torque vanishes. Another increase in $\tilde{V}$ 
leads to negative torques, favoring orientation of the colloid parallel to the 
surface.

\begin{figure}[!th]
\begin{center}
\includegraphics[width=0.45\textwidth,bb=95 230 485 540,clip]{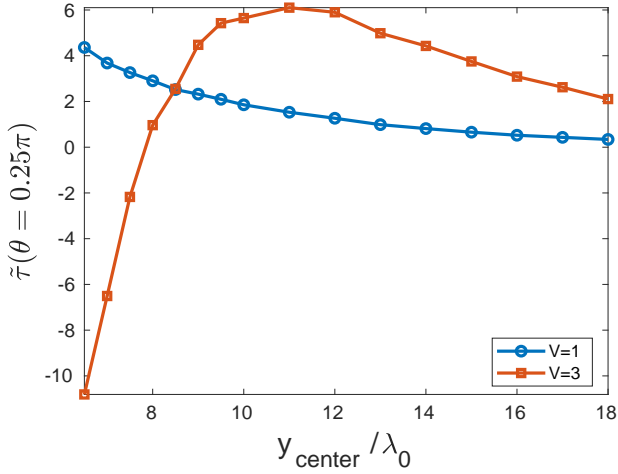}
\caption{Dimensionless torque $\tilde{{\bf \tau}}$ vs
increasing values of colloid's distance to the surface $y_{\rm center}$ at a 
fixed tilt angle $\theta=45^\circ$. At small potentials, the torque is positive 
and decreasing with $y_{\rm center}$, favoring orientation of the colloid in 
the direction perpendicular to the surface ($\tilde{V}=1$, blue curve). At 
larger potentials ($\tilde{V}=3$, red), close to the surface the torque acts to 
orient the colloid parallel to the surface while at larger distances it orients 
the colloid perpendicular to the surface.
$a=6\lambda_0$ and $a/b=3.2$. (For interpretation of the references to colour in 
this figure legend, the reader is referred to the web version of this article.)
}
\label{fig4}
\end{center}
\end{figure}

One can also look at the variation of the torque with colloid distance from the 
wall. In Fig. \ref{fig4} we show the torque for a colloid tilted at 
$\theta=45^\circ$ with respect to the $y$ axis vs $y_{\rm center}$. If the 
surface potential is small, $\tilde{V}=1$, blue curve, the torque is positive, 
leading to  CCW rotation. The torque decreases with increasing distance. 
However, the behavior is very different when the potential is large. The red 
curve ($\tilde{V}=3$) starts from negative values, where the torque 
tends to orient the colloid in the CW direction. As the distance increases the 
torque increases, and becomes positive at a large enough $y_{\rm center}$. At 
even larger values it decreases again and tends to zero as $y_{\rm 
center}\to\infty$.

\begin{figure}[!th]
\begin{center}
\includegraphics[width=0.48\textwidth,bb=20 270 720 
480,clip]{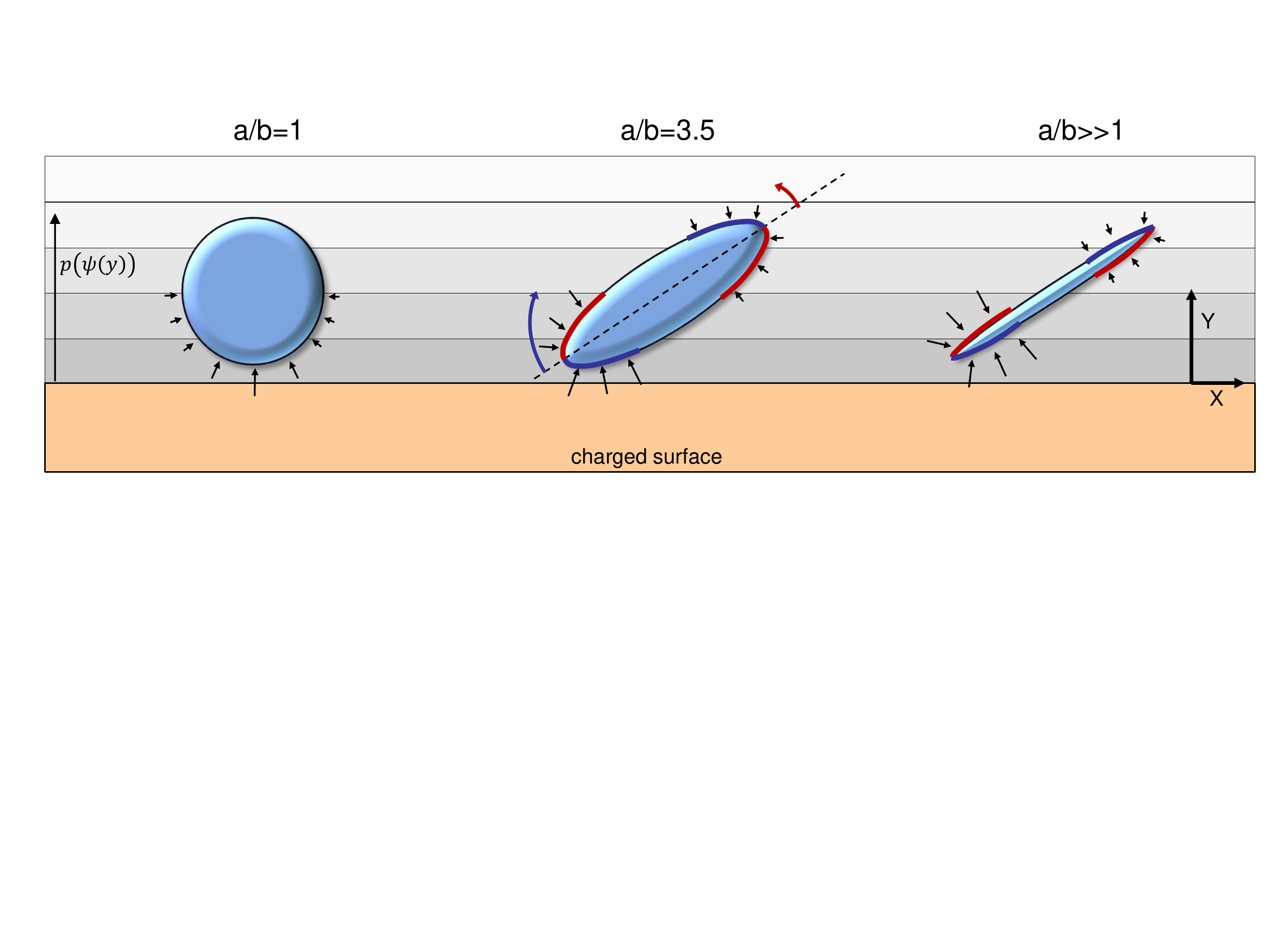}
\caption{Illustration of forces acting on colloids with different 
aspect ratios in a polar solvent near a charged surface. Horizontal grey shades 
grossly correspond to the ideal gas pressure of ions varying nonlinearly in the 
$y$-direction. The forces exerted on the colloid by the ion gas are shown by 
arrows; longer arrows close to the surface mean stronger forces. Left: for the
spherical colloid forces point to the center and the torque vanishes. Center: 
for an elongated colloid the force for CW (blue) and CCW (red) rotations do not 
balance -- the force at the left-bottom tip for CW rotation is stronger than 
the CCW rotation force at the top-right tip. The torque exerted by the ions 
tends to zero again for a needle-like colloid (right, $a/b\gg 1$) because the 
pressure at the two sides of the colloid (blue and red segments) becomes the 
same when its width tends to zero. (For interpretation of the references to 
colour in this figure legend, the reader is referred to the web version of this 
article.)
}
\label{fig_illustration_orientation}
\end{center}
\end{figure}

In Figs. \ref{fig2}, \ref{fig3}, and \ref{fig4} a transition from 
the classical colloidal orientation perpendicular to the surface to an 
unusual orientation parallel to the surface is clearly seen.
What is the origin of this unexpected transition? Close inspection of the 
forces shows that the transition is due to 
the ideal-gas pressure $p_0$. Figure \ref{fig_illustration_orientation} 
is an illustration of the ideal-gas forces acting on 
colloids with different aspect ratios. The forces exerted by the gas of ions are 
shown by arrows (long arrow denotes strong force). The horizontal grey level 
shades denote the ion gas pressure $p_0\sim 2n_0\cosh(\tilde{\psi})$.
On the left, the forces acting on the spherical colloid 
($a/b=1$) point towards its center and the 
torque vanishes. The colloid with $a/b=3.5$ (center) experiences nonuniform 
forces: near the left-bottom tip, the forces for CW rotation (blue segment) are 
stronger than the forces in the CCW direction (red segment), leading to a net 
torque in the CW direction. Near the opposite tip the torques are 
reversed. 
However, since the pressure decays rapidly with $y$ these torques are
negligible and the total torque is CW. In the limit of a slender colloid 
with large aspect ratio $a/b\gg 1$ (right), the blue and red 
segments experience nearly the same force and the torque by the ion gas tends 
to zero.

\begin{figure}[!th]
\begin{center}
\includegraphics[width=0.4\textwidth,bb=95 240 480 540,clip]{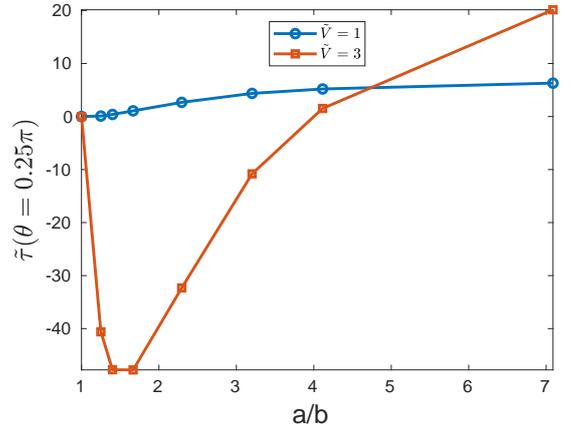}
\caption{Dimensionless torque $\tilde{{\bf \tau}}$ at fixed tilt angle vs 
increasing colloid aspect ratio values $a/b$. At small potentials 
($\tilde{V}=1$, blue curve) the torque is positive, favoring rotation of the 
colloid perpendicular to 
the surface. The torque increases with $a/b$ as the colloid is more elongated. 
For larger potentials ($\tilde{V}=3$, red curve) the torque decreases 
rapidly to a negative minimum, and then increases with increasing aspect ratio 
$a/b$. It becomes positive again for needle-like colloids.
(For interpretation of the references to colour in this figure legend, the 
reader is referred to the web version of this article.)
}
\label{fig6}
\end{center}
\end{figure}

From the above discussion it is clear that there is a finite value of $a/b$ 
that maximizes the torque for CW rotation.
Indeed this behavior is shown in Fig. \ref{fig6}, where 
the torque is plotted vs colloid aspect ratio $a/b$. For small voltages, the 
ion gas pressure effect is not strong enough to overcome the ``classical'' 
behavior and the torque orients the colloid perpendicular to the surface. If the 
voltage is strong enough, $\tilde{V}=3$, the most negative torque occurs when 
$a/b\approx 1.5$. As $a/b$ increases the torque increases from its minimal value 
to higher values, until the classical behavior overcomes and the torque becomes 
positive. 

%


\section{Conclusions}

We show that colloidal orientation near charged 
surfaces is bi-stable. The classical behavior where the colloid aligns with its 
long axis parallel to the field occurs at large colloid-surface distances or 
small surface potentials. 
The orientation with the long axis parallel to the 
surface and perpendicular to the field is uncommon. To the best of our 
knowledge only 
one work reports a similar transition -- Buyukdagli and Podgornik 
\cite{rudi_pre_2019}. These authors looked at correlation corrections to the 
mean-field theory to describe a charged 
rod near charged membrane in the weak and intermediate charge regimes. 
In our Poisson-Boltzmann theory the transition is first-order.
The above analysis was carried out 
in the regime where the colloid size is neither much larger than Debye's 
length $\lambda_D$ (as is typically in electroosmosis 
\cite{holm_langmuir_2014}) where gradients are 
localized at the colloid's surface, nor is it much smaller than $\lambda_D$, 
where the field is essentially uniform.

The driving force for the orientation parallel to the surface is the ideal gas 
pressure of dissolved ions. This pressure varies nonlinearly with $y$ and this 
is essential for the rotation of the colloid. This is different 
from, e.g., a 
solid body submerged in a liquid under gravity on Earth. As is well known such 
a body feels an upward or downward force depending on its density 
and proportional to the volume 
and the total torque vanishes. But this is true only when the 
pressure 
varies linearly with the depth $y$. The scaling of the effective 
potenail, $U_{\rm eff}/k_BT\sim n_0a^3$, where $n_0$ is the ion number density 
in solution, highlights the importance of the ion gas pressure and is large 
because $U_{\rm eff}$ is effectively the integral of the surface torques, 
namely $U_{\rm eff}\sim\veps E^2a^3$, and $E\sim k_BT/e\lambda_D$.

Our findings suggest a novel way to control the 
orientation of a colloidal suspension near a wall by a simple modification of 
the surface potential. For practical use one may want to cover the electrode 
with an insulating layer as is commonly done in electrowetting on dielectrics. 
For the best results it would be desired to use colloids with an 
optimal shape. 
In this work we considered spheroidal particles and the optimality is in the 
aspect ratio $a/b$, Fig. \ref{fig6}. Symmetry-breaking can arise from 
surface inhomogeneities and not necessarily from asymmetric shapes.  
Chemical composition gradients, e.g. metallic coating of parts of the colloid, 
imply asymmetric distribution of the field surrounding the colloid, and will 
lead to non-trivial torque on the colloid \cite{bazant_prl_2008}.

The orientational transition can have rheological 
consequences as well: we speculate that when a suspension is 
forced to flow parallel to the surface, the flow regime and effective viscosity 
would be very different in the two colloidal states, $\theta=0$ 
or $\theta=90^\circ$.

We neglected van der Waals and Casimir interactions but such 
forces acting via 
the nonhomogeneous electrolytes may have significant implications 
\cite{rudi_jcp_2016,rudi_rmp_2016}.
This work assumed a static colloid at a given fixed position $y_{\rm 
center}$ and tilt angle $\theta$. The force ${\bf F}$ and torque 
$\tau$ both depend on $\theta$. It would be interesting to 
simulate the 
total undulatory ``swimming'' motion of such a colloid when
translation and rotation are coupled, and the transition between the $\theta=0$ 
and $\theta=\pi/2$ orientations. Here hydrodynamic and van der 
Waals interactions between two or 
more colloids will play a vital role due to the presence of the wall 
\cite{squires_prl_2000,dufresne_prl_2000}.

\noindent
{\bf Acknowledgments}

This work was supported by the Israel Science Foundation Grant No. 56/14. I 
thank Antonio Ramos for useful correspondence and comments.

\noindent
{\bf Appendix A: Supplementary material} Numerical data and code will be 
supplied upon request.

\bibliography{mybibfile}

\end{document}